\documentclass[aps,twocolumn,prb,footinbib,showpacs,superscriptaddress,
preprintnumbers,amsmath,amssymb]{revtex4-1}

\usepackage{graphicx}
\usepackage{subfigure}
\usepackage{epsfig}
\usepackage{dcolumn}
\usepackage{bm} 
\usepackage{color}

\begin{document}


\title{Thermal rectification in a double quantum dots system with polaron effect}

\author{Gaomin Tang}
\affiliation{Department of Physics and the Center of Theoretical and Computational Physics, The University of Hong Kong, Hong Kong, China}
\affiliation{Department of Physics, National University of Singapore, Singapore 117551, Republic of Singapore}
\author{Lei Zhang}
\email{zhanglei@sxu.edu.cn}
\affiliation{State Key Laboratory of Quantum Optics and Quantum Optics Devices, Institute of Laser Spectroscopy, Shanxi University, Taiyuan 030006, China}
\affiliation{Department of Physics and the Center of Theoretical and Computational Physics, The University of Hong Kong, Hong Kong, China}
\affiliation{Collaborative Innovation Center of Extreme Optics, Shanxi University, Taiyuan 030006, China}
\author{Jian Wang}
\email{jianwang@hku.hk}
\affiliation{Department of Physics and the Center of Theoretical and Computational Physics, The University of Hong Kong, Hong Kong, China}

\date{\today}

\begin{abstract}
We investigate the rectification of heat current carried by electrons through a double quantum dot (DQD) system under a temperature bias. The DQD can be realized by molecules such as suspended carbon nanotube and be described by the Anderson-Holstein model in presence of electron-phonon interaction. 
Strong electron-phonon interaction can lead to formation of polaronic states in which electronic states are dressed by phonon cloud.
Dressed tunneling approximation (DTA), which is nonperturbative in dealing with strong electron-phonon interaction, is employed to obtain the heat current expression. In DTA, self-energies are dressed by phonon cloud operator and are temperature dependent.
The temperature dependency of imaginary part of dressed retarded self-energy gives rise to the asymmetry of the system and is the necessary condition of thermal rectification. On top of this, one can either tune DQD effective energy levels such that $|\bar{\epsilon}_1|\neq |\bar{\epsilon}_2|$ or have asymmetric dot-lead couplings to achieve thermal rectification. We numerically find that increasing electron-phonon coupling and reducing inter dot coupling can both improve thermal rectification effect, while the electronic heat current is reduced. 
\end{abstract}

\maketitle

\section{Introduction}\label{sec1}
	Understanding and controlling heat flow become essential for a variety of applications in heating, refrigeration, heat-assist information storage, and energy conversion in modern times \cite{RMP_Li, thermal_rev, harvester}. In analogy to electric diodes as the primary building block in the electronics industry, thermal diode or rectifier becomes crucial in managing heat flow by changing heat current magnitude with the reversal of temperature bias.
Thermal rectification has been reported in various systems such as magnonic junction \cite{magnon_diode}, spin Seebeck engine \cite{Jie1, gm5}, topological insulator-superconductor junction \cite{TIS_diode}, normal metal-superconductor junction \cite{NS_diode0, NS_diode1, NIS_diode2}, near field raditative heat transfer \cite{NFRHT_diode}, geometric asymmetric structures \cite{shape1, shape2, shape3, shape4}, etc..
Heat current can be transferred via different types of energy carriers \cite{Segal0, Segal1}, such as magnons \cite{magnon_diode, Jie1, gm5}, electrons \cite{TIS_diode, NS_diode0, NS_diode1, NIS_diode2, lei, yu}, photons \cite{NFRHT_diode, NFRHT_diode2}, phonons \cite{phonon_diode}, acoustic wave \cite{acoustic1, acoustic2}, and etc..
Regardless of the type of the junctions and energy carriers, the key ingredient required for achieving thermal rectification is structural asymmetry \cite{asymmetry}.


	Over the past decades, rapid experimental development of nanotechnology enables us to fabricate the single-molecule junctions \cite{molecule0, molecule1, molecule2}. The elastic mechanical deformation caused by charging of the molecule can give rise to electron-phonon interaction. Single-molecule junction could be modeled and measured as a quantum dot or serial double quantum dots (DQD) described by the Anderson-Holstein model \cite{Holstein, Mahan} connected to two leads. Suspended carbon nanotubes (CNTs), which are free to oscillate due to high Q factors and stiffness, become very favorable in experiments \cite{CNT}. Recently, it has been reported that electron-phonon coupling strength of a suspended CNT can be tailored \cite{tailor}.
The strong electron-phonon coupling in molecular junctions can lead to the polaronic regime in which electronic states are dressed by phonon cloud \cite{ep_polaron1, ep_polaron2}. The polaronic effect exhibits novel transport properties, such as negative differential conductance \cite{molecule2, NDC1, NDC2}, phonon-assisted current steps \cite{NDC2, BDong1, DTA, BDong2, FC_blockade1, FC_blockade2}, Franck-Condon blockade \cite{FC_blockade1,FC_blockade2, Patrick1}, sign change in the shot noise correction \cite{Schmidt}, and etc.. Rectification effect of electric current in a single molecular dimer, which is modeled as a DQD with strong polaron effect, has been reported \cite{molecule_dimer}, while the thermal rectification counterpart is yet to be revealed.

	In this work, we report the rectification effect of electronic heat current through a DQD system in polaronic regime under a temperature bias. A nonperturbative approach, i.e., the dressed tunneling approximation (DTA) \cite{BDong1, DTA, BDong2} is employed to deal with strong electron-phonon interaction. Self-energies dressed by phonon cloud operator due to electron-phonon couplings are introduced. The expression of electronic heat current is obtained from the equation of motion technique together with DTA. We find that the necessary condition to realize the rectification effect in the DQD system is the temperature dependency of dressed retarded self-energies. In addition, one should either tune the DQD levels $\bar{\epsilon}_\nu$ satisfying $|\bar{\epsilon}_1| \neq |\bar{\epsilon}_2|$ or set the dot-lead couplings unequal to achieve thermal rectification.
We show the behaviors of dressed retarded self-energy with respect to temperatures and electron-phonon coupling constant $g$. We find that increasing $g$ and decreasing inter dot coupling can improve rectification effect, while heat current is reduced.

	The rest of the article is organized as follows. In Sec. II, the model Hamiltonian of the DQD molecular junction is introduced and the expression of electronic heat current is given in terms of nonequilibrium Green's function (NEGF) formalism. In Sec. III, since the temperature dependency of dressed retarded self-energy is crucial to achieve thermal rectification, we first show its behaviours with respect to different temperatures and electron-phonon coupling. Then we investigate the heat current rectification effect by changing system parameters. Finally, a brief conclusion is drawn in Sec. IV.

\section{Model and theoretical formalism}
\begin{figure}
  \includegraphics[width=2.7in]{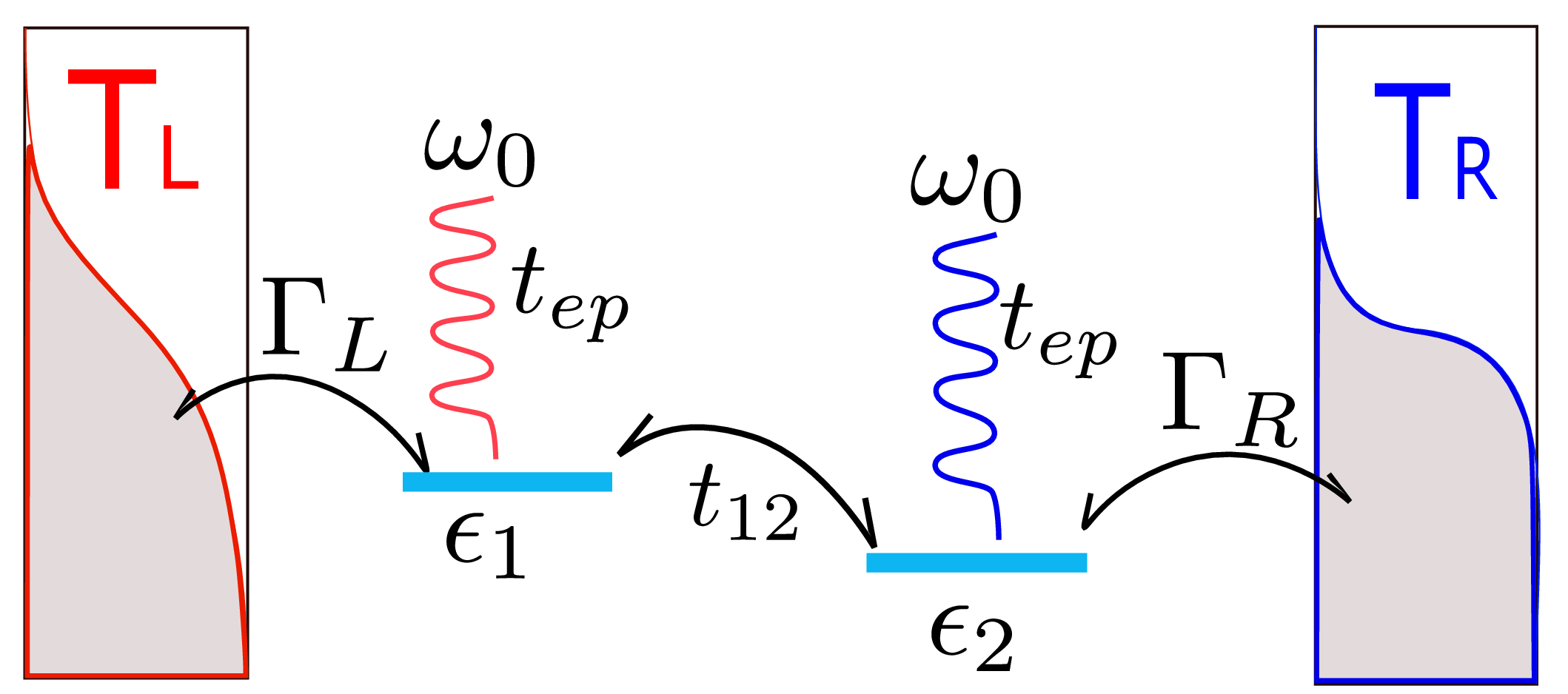} \\
  \caption{(Color online) Sketch of a DQD coupled to the left and right lead with temperature $T_L$ and $T_R$. Each quantum dot interacts with a local vibrational mode, which has the same temperature with the corresponding lead. The central DQD can be described by Anderson-Holstein model.  }
  \label{fig1}
\end{figure}

We consider our setup as a serial DQD system made of molecules connected to its left and right lead with different temperatures (see Fig.~\ref{fig1}).
The DQD can be realized using CNT, with the central part being fixed and each lateral part suspended as the quantum dot \cite{NDC1}. Each quantum dot shall be close to the lead that it connects with, so that the localized vibrational mode of each quantum dot bears the same temperature with corresponding lead. The hopping between the two quantum dots can be tuned using a gate voltage applied on the central fixed part.
The central DQD can be described by Anderson-Holstein model \cite{Holstein, Mahan} and its Hamiltonian is expressed as,
\begin{align}
H_{DQD} = &\sum_{\nu=1,2} \big[\epsilon_\nu \hat{n}_\nu + t_{ep\nu}(a_\nu^\dag +a_\nu) \hat{n}_\nu + \omega_0 a_\nu^\dag a_\nu \big]  \notag \\
&+ t_{12} (d_1^\dag d_2 + d_2^\dag d_1) .
\end{align}
Here, $\epsilon_\nu$ is the bare electronic energy level of the site $\nu$ quantum dot,  $\omega_0$ is the frequency of the localized phonon, $d_\nu^\dag$ ($a_\nu^\dag$) denotes the electron (phonon) creation operator, and the occupation operator is $\hat{n}_\nu = d_\nu^\dag d_\nu$. $t_{12}$ is the inter dot hopping amplitude. We assume that electron-phonon couplings in DQD are the same constant with $t_{ep1}=t_{ep2}=t_{ep}$.
The total Hamiltonian reads as
\begin{equation}
H = H_{DQD} + \sum_\alpha H_{\alpha} + H_T ,
\end{equation}
with the Hamiltonians of the leads
\begin{equation}
H_{\alpha} = \sum_{k;\alpha} \epsilon_{k\alpha} c_{k\alpha}^\dag c_{k\alpha} ,
\end{equation}
where $c^\dagger_{k\alpha}$ creates an electron in lead $\alpha$ and the tunneling Hamiltonian between the dots and the leads,
\begin{equation}
H_T = \sum_{k} \big(t_{kL} c_{kL}^\dag d_1 + t_{kR} c_{kR}^\dag d_2 + {\rm H.c.} \big) .
\end{equation}
The indices $k\alpha = kL, kR$ are used to label the different states in the left and right leads, and $t_{k\alpha}$ is the tunneling amplitude between quantum dot and state $k$ in lead $\alpha$. The tunneling rates of both leads are assumed to be independent of energy (wide band limit) as
\begin{equation} \label{linewidth}
{\bf \Gamma}_\alpha(\omega) ={\rm Im} \sum_k \frac{|t_{k\alpha}|^2}{\omega-\epsilon_{k\alpha}-i0^+} = \Gamma_\alpha ,
\end{equation}
and one can denote $\Gamma = \Gamma_L+\Gamma_R$. The leads chemical potentials are both set to zero $\mu_L=\mu_R=0$, and a temperature bias is applied at two leads with $\Delta T = T_L-T_R$ to induce electronic transport.

Applying the Lang-Firsov unitary transformation \cite{Lang-Firsov} given by
\begin{equation}
 \bar{H} = S H S^\dag, \ \ S=\exp\left[ g \sum_\nu{ \hat{n}_\nu (a_\nu^\dag -a_\nu)} \right], \ \
 g = \frac{t_{ep}}{\omega_0},
\end{equation}
one can eliminate the electron-phonon coupling term, and get the central DQD Hamiltonian as,
\begin{equation}
\bar{H}_{DQD} =\sum_{\nu=1,2} \big[ \bar{\epsilon}_\nu \hat{n}_\nu +\omega_0 a_\nu^\dag a_\nu \big] + t_{12} (d_1^\dag d_2 + d_2^\dag d_1),
\end{equation}
where the effective bare QD electronic energies are changed to be $\bar{\epsilon}_\nu=\epsilon_\nu - g^2\omega_0$. Throughout this work, the effective bare QD electron energies are termed as QD electron energies for convenience.
The tunneling Hamiltonian is transformed to be
\begin{equation}
\bar{H}_T = \sum_{k} (t_{kL} c_{kL}^\dag X_1 d_1 + t_{kR} c_{kR}^\dag X_2 d_2 + {\rm H.c.} ) ,
\end{equation}
with the phonon cloud operator $X_\nu = \exp[g(a_\nu-a_\nu^\dag)]$, while the Hamiltonian of the uncoupled leads remains unchanged.

	The energy current carried by electrons through lead $\alpha$ is defined as the time evolution of the Hamiltonian of lead $\alpha$ with the form \cite{gm4, gm3} (in natural units, $\hbar = k_B =e = m_e =1$),
\begin{equation}
I^E_\alpha =-\left\langle \dot{\hat{H}}_\alpha \right\rangle
=-i\left\langle [\hat{H}_\alpha, \bar{H}]\right\rangle.
\end{equation}
Since no voltage bias is applied, the energy current is the same as the heat current, and the electronic heat current through the left lead reads as \cite{gm3}
\begin{equation}
I_L^h = \sum_k \left[ i\epsilon_{kL} t_{kL} \langle c_{kL}^\dag (t) X_1(t) d_1(t) \rangle + {\rm H.c.} \right].
\end{equation}
One can define the following contour ordered Green's functions,
\begin{align}
\widetilde{G}_{1,kL}(\tau_1,\tau_2) =& -i \langle {\cal T}_C c_{kL}^\dag(\tau_2) X_1(\tau_2) d_1(\tau_1) \rangle ; \label{GreenFunction} \\
\widetilde{G}_{2,kR}(\tau_1,\tau_2) =& -i \langle {\cal T}_C c_{kR}^\dag(\tau_2) X_2(\tau_2) d_2(\tau_1) \rangle ,
\end{align}
where ${\cal T}_C$ is the time ordering operator in the Keldsyh contour. From now on, the time on the contour is denoted using Greek letters, and real time using Latin letters. This implies that $\widetilde{G}_{1,kL}(\tau_1,\tau_2)$ and $\widetilde{G}_{2,kR}(\tau_1,\tau_2)$ are two-by-two matrices with their entries to be $\widetilde{G}^{ab}_{1,kL}(t_1,t_2)$ and $\widetilde{G}^{ab}_{2,kR}(t_1,t_2)$, respectively. Here $a,b=+,-$ denote the different branches of the contour.
The lesser and greater Green's functions defined in Eq.~\eqref{GreenFunction} read, respectively, as
\begin{align}
\widetilde{G}_{1,kL}^{+-} (t_1,t_2) &= -i \langle c_{kL}^\dag(t_2) X_1(t_2) d_1(t_1) \rangle ; \\
\widetilde{G}_{1,kL}^{-+} (t_1,t_2) &= -i \langle d_1^\dag(t_1) X_1^\dag(t_2) c_{kL}(t_2) \rangle .
\end{align}
Then heat current is given by
\begin{equation} \label{I_Lh}
I_L^h(t) = \sum_k \epsilon_{kL} t_{kL} \left[ \widetilde{G}_{1,kL}^{-+} (t,t) - \widetilde{G}_{1,kL}^{+-} (t,t) \right] .
\end{equation}

	Having defined the Green's functions and heat current carried by electrons, we next employ equation of motion and  dressed tunneling approximation (DTA) to get the final heat current expression in terms of Fermi distribution functions and transmission coefficient function.
The equation of motion of the three point Green function on the contour $\left\langle {\cal T}_C c_{kL}^\dag(\tau_2) X_1(\tau') d_1(\tau_1)\right\rangle$ is
\begin{align}
&\left(i\frac{\partial}{\partial \tau_2} -\epsilon_{kL}\right)\left\langle {\cal T}_C c_{kL}^\dag(\tau_2) X_1(\tau') d_1(\tau_1) \right\rangle  \notag \\
&=t_{kL}^* \left\langle {\cal T}_C d_1^\dag(\tau_2) X_1^\dag(\tau_2) X_1(\tau') d_1(\tau_1) \right\rangle .
\end{align}
The differential form could be written in an integral form \cite{Haug} as
\begin{align} \label{integral}
&\left\langle {\cal T}_C c_{kL}^\dag(\tau_2) X_1(\tau') d_1(\tau_1) \right\rangle  \notag \\
&= \int_C d\tau \left\langle {\cal T}_C d_1^\dag(\tau) X_1^\dag(\tau) X_1(\tau') d_1(\tau_1) \right\rangle t_{kL}^* g_{kL}(\tau,\tau_2) ,
\end{align}
where $g_{kL}(\tau,\tau_2)$ is the free electronic Green's function of the state $k$ in the left lead.
One can define the DQD Green's function on the Keldysh contour as,
\begin{equation} \label{Gnu}
G_{\nu\nu'}(\tau_1,\tau_2)= -i\langle {\cal T}_C d_{\nu'}^\dag (\tau_2) d_\nu(\tau_1) \rangle ,
\end{equation}
with $\nu,\nu'=1,2$ labelling quantum dots.


The electron-phonon interaction is usually treated perturbatively for weak electron-phonon coupling \cite{weak}. However, in the strong electron-phonon coupling regime, the traditional perturbation technique fails and a non-perturbative approximation is needed. With the strong electron-phonon coupling and weak lead-dot coupling, the lifetime of the electronic states in DQD is much larger than that in the leads, DTA is suitable and can cope with the pathological features of the single particle approximation at low frequencies and polaron tunneling approximation at high frequencies \cite{DTA}. Under DTA, one has the decoupling \cite{DTA}
\begin{align} \label{decoupling}
&\left\langle {\cal T}_C d_1^\dag(\tau) X_1^\dag(\tau) X_1(\tau') d_1(\tau_1) \right\rangle
\notag \\
\simeq &\left\langle {\cal T}_C X_1^\dag(\tau) X_1(\tau') \right\rangle \left\langle {\cal T}_C d_1^\dag(\tau) d_1(\tau_1) \right\rangle  \notag \\
= & i\Lambda_1(\tau',\tau) G_{11}(\tau_1,\tau) ,
\end{align}
and similarly
\begin{equation}
\left\langle {\cal T}_C d_2^\dag(\tau) X_2^\dag(\tau) X_2(\tau') d_2(\tau_1) \right\rangle
= i\Lambda_2(\tau',\tau) G_{22}(\tau_1,\tau) ,
\end{equation}
with $\Lambda_\nu(\tau',\tau) = \left\langle {\cal T}_C X_\nu^\dag(\tau) X_\nu(\tau') \right\rangle$ being the phonon cloud propagator. From Eqs.~\eqref{GreenFunction}, \eqref{integral}, and \eqref{decoupling}, we have
\begin{align}  \label{G1kL}
\widetilde{G}_{1,kL}(\tau_1,\tau_2) t_{kL} = \int_C d\tau G_{11}(\tau_1,\tau) t_{kL}^* g_{kL}(\tau,\tau_2) t_{kL}\Lambda_1(\tau_2,\tau) .
\end{align}
The above equation is expressed in Keldysh space and subscript `$C$' denotes the time integration in Keldysh contour. Performing the summation over states $k$, we obtain 
\begin{equation} 
\sum_k\widetilde{G}_{1,kL}(\tau_1,\tau_2) t_{kL} = \int_C d\tau G_{11}(\tau_1,\tau) \Sigma_L(\tau,\tau_2) \Lambda_1(\tau_2,\tau) ,
\end{equation}
where the self-energy is expressed as
\begin{equation}
\Sigma_L(\tau,\tau_2)= \sum_{k} t_{kL}^* g_{kL}(\tau,\tau_2) t_{kL}, 
\end{equation}
with the Keldysh components
\begin{equation}
\Sigma_L^{ab}(t,t_2)= ab\sum_{k} t_{kL}^* g_{kL}^{ab}(t,t_2) t_{kL} .
\end{equation}
A similar expression for $\sum_k\widetilde{G}_{2,kR}(\tau_1,\tau_2) t_{kR}$ can be obtained as well. 


For convenience, we use the lead index $\alpha$ instead of $\nu$ as the sub-index for the phonon cloud operator so that $\Lambda_L^{ab}(t_1,t_2)\equiv \Lambda_1^{ab}(t_1,t_2)$ and $\Lambda_R^{ab}(t_1,t_2)\equiv \Lambda_2^{ab}(t_1,t_2)$.
The self-energies dressed by the phonon cloud propagator under the DTA are then expressed as,
\begin{equation} \label{dress}
\Sigma_{d\alpha}^{ab} (t_1,t_2) = \Sigma_\alpha^{ab} (t_1,t_2) \Lambda_\alpha^{ab}(t_1,t_2) .
\end{equation}
The lesser and greater phonon cloud operator $\Lambda_\alpha^{ab}(t_1,t_2)$ are given by \cite{Mahan},
\begin{equation} \label{phonon}
\Lambda_\alpha^{+-}(t_1,t_2) = \left[ \Lambda_\alpha^{-+}(t_1,t_2) \right]^*
=\sum_{m=-\infty}^{\infty} A_{\alpha m} e^{i m\omega_0 (t_1-t_2)},
\end{equation}
with
\begin{equation}
A_{\alpha m} = e^{-g^2(2n_{\alpha}+1)} e^{m\beta_\alpha\omega_0/2} I_m\left( 2g^2\sqrt{n_{\alpha}(1+n_{\alpha})}\right),
\end{equation}
$I_m$ the modified Bessel function of the first kind, and Bose factor $n_{\alpha}=1/(e^{\beta_\alpha\omega_0}-1)$, $\beta_\alpha=1/k_B T_\alpha$. The remaining time-ordered and anti-time-ordered components could be calculated through the relations,
\begin{align}
&\Lambda_\alpha^{++}(t_1,t_2)=\theta(t_1-t_2)\Lambda_\alpha^{-+}(t_1,t_2)
+\theta(t_2-t_1)\Lambda_\alpha^{+-}(t_1,t_2), \notag \\
&\Lambda_\alpha^{--}(t_1,t_2)=\theta(t_2-t_1)\Lambda_\alpha^{-+}(t_1,t_2)
+\theta(t_1-t_2)\Lambda_\alpha^{+-}(t_1,t_2).
\end{align}

	Taking the time derivative of the DQD NEGF $G_{\nu\nu'}(\tau_1,\tau_2)$ in Eq.~\eqref{Gnu} \cite{Haug} can give us the contour ordered Dyson equation of the system
\begin{align}
&{\bf G}(\tau_1,\tau_2)= {\bf g}(\tau_1,\tau_2) +  \notag \\
&\int\int_C d\tau d\tau' {\bf g}(\tau_1,\tau) \left[ {\bf \Sigma}_d(\tau,\tau') + {\bf t}\delta(\tau-\tau') \right] {\bf G} (\tau',\tau_2) ,
\end{align}
with
\begin{align*}
{\bf G} = \begin{pmatrix}
G_{11} & G_{12} \\ G_{21} & G_{22}
\end{pmatrix} ,  \qquad
&{\bf g} = \begin{pmatrix}
g_1 &   \\   & g_2
\end{pmatrix},   \\
{\bf \Sigma}_d = \begin{pmatrix}
\Sigma_{dL} &  \\   &  \Sigma_{dR}
\end{pmatrix} ,  \qquad
&{\bf t} = \begin{pmatrix}
 & t_{12}  \\  t_{21}  &
\end{pmatrix}.
\end{align*}
where $g_\nu$ is the free DQD Green's function of site $\nu$ without coupling to the leads.
In the wide band limit and through Fourier transformation of Eq.~\eqref{dress}, the dressed lesser and greater self-energy in the energy domain are given by
\begin{align}
\Sigma_{d\alpha}^{+-}(\omega) &=-\sum_{m} A_{\alpha m} i\Gamma_\alpha f_{\alpha+m} , \\
\Sigma_{d\alpha}^{-+}(\omega) &=-\sum_{m} A_{\alpha m} i\Gamma_\alpha \left[ f_{\alpha-m}-1 \right] ,
\end{align} 
where $f_{\alpha+m}=1/[\exp(\beta_\alpha (\omega+m\omega_0-\mu_\alpha))+1]$.
The dressed retarded self-energy in energy domain is obtained throught the Fourier transformation of the time domain counterpart $\Sigma_d^r(t_1,t_2)=\theta(t_1-t_2)\left[\Sigma_d^{+-}(t_1,t_2)-\Sigma_d^{-+}(t_1,t_2)\right]$, so that, \cite{BDong1, DTA, BDong2}
\begin{equation}  \label{dressed_r}
\Sigma_{d\alpha}^{r}(\omega) =\sum_{m} A_{\alpha m}\int \frac{dE}{2\pi} \frac{\Gamma_\alpha \left[1-f_{\alpha-m}(E)+f_{\alpha+m}(E)\right]}{\omega-E+i0^+} .
\end{equation}
The real and imaginary part of the dressed retarded self-energy can be obtained using Plemelj formula $1/(E\pm i0^+)=P(1/E)\mp i\pi\delta(E)$ which facilitates the numerical calculation as well. We can directly prove that the real part and imaginary part are odd and even function with respect to chemical potential, respectively \cite{gm4, BDong1} with the expressions,
\begin{align} \label{Sr}
{\rm Re}\left[ \Sigma_{d\alpha}^{r}(\mu_\alpha+\omega) \right] &= -{\rm Re}\left[ \Sigma_{d\alpha}^{r}(\mu_\alpha-\omega) \right] , \notag \\
{\rm Im}\left[ \Sigma_{d\alpha}^{r}(\mu_\alpha+\omega) \right] &= {\rm Im}\left[ \Sigma_{d\alpha}^{r}(\mu_\alpha-\omega) \right] .
\end{align}
Since the chemical potential of both leads are set to zero, the real and imaginary part of dressed retarded self-energy are even and odd function of energy. By performing Keldysh rotation \cite{Ka2}, one can get the Dyson equation for the retarded Green's function in the energy domain as
\begin{equation} \label{Dyson_r}
{\bf G}^r = {\bf g}^r + {\bf g}^r \left( {\bf \Sigma}_d^r + {\bf t} \right) {\bf G}^r .
\end{equation}
Similar equation applies for the advanced Green's function ${\bf G}^a$. The lesser and greater Green's functions are given by the following Keldysh equation
\begin{align}
{\bf G}^{+-} &=  {\bf G}^r {\bf \Sigma}_d^{+-} {\bf G}^a , \\
{\bf G}^{-+} &=  {\bf G}^r {\bf \Sigma}_d^{-+} {\bf G}^a .
\end{align}

	From Eqs. \eqref{I_Lh} and \eqref{G1kL}, heat current can be expressed as \cite{gm3},
\begin{equation}
I_L^h = \int dt' \left[ G_{11}^{+-}(t,t')\breve{\Sigma}_{dL}^{-+}(t',t) - G_{11}^{-+}(t,t')\breve{\Sigma}_{dL}^{+-}(t',t) \right] ,
\end{equation}
where
\begin{equation}
\breve{\Sigma}_{dL}^{+-}(t',t)
=-\Lambda_L^{+-}(t'-t) \sum_k \epsilon_{kL} t_{kL}^* g_{kL}^{+-}(t'-t) t_{kL} .
\end{equation}
and similarly for $\breve{\Sigma}_{dR}^{-+}(t',t)$.
	In the long time limit, one can prove the heat current conservation law $I_L^h=-I_R^h \equiv I_h$, and the current expression could be expressed by an integral in the energy domain as
\begin{align}
&I_h = \int \frac{d\omega}{2\pi} \hbar\omega \left[ G_{11}^{+-}(\omega) \Sigma_{dL}^{-+}(\omega) - G_{11}^{-+}(\omega) \Sigma_{dL}^{+-}(\omega) \right] \notag \\
&= \int \frac{d\omega}{2\pi} \hbar\omega |G_{12}^r(\omega)|^2 [\Sigma_{dL}^{-+}(\omega) \Sigma_{dR}^{+-}(\omega) - \Sigma_{dL}^{+-}(\omega) \Sigma_{dR}^{-+}(\omega)] .
\end{align}
Further simplification enables us to arrive at the heat current expression
\begin{widetext}
\begin{equation}
I_h=\int \frac{d\omega}{2\pi} \hbar\omega T(\omega) \sum_{mn} A_{Lm} A_{Rn} \left[f_{L+m}(1-f_{R-n})- f_{R+n}(1-f_{L-m}) \right].
\end{equation}
\end{widetext}
In this expression $T(\omega)$ is the transmission coefficient with the form,
\begin{equation}
T(\omega) = |G_{12}^r(\omega)|^2\Gamma_L\Gamma_R .
\end{equation}
The expression of $G_{12}^r(\omega)$ is obtained from Eq.~\eqref{Dyson_r},
\begin{equation} \label{G12}
G_{12}^r(\omega) = \frac{t_{12}}{ [\omega-\bar{\epsilon}_1-\Sigma_{dL}^r(\omega)][\omega-\bar{\epsilon}_2-\Sigma_{dR}^r(\omega)]-t_{12}^2 }.
\end{equation}
We can see that DTA provides a clear physical picture for phonon-assisted electronic heat flow. An electron coming from the left lead with energy $\omega$ absorbs (emits) $m>0$ ($m<0$) vibrational quanta and tunnels into the DQD, and then flows into the right lead by emitting (absorbing) $n<0$ ($n>0$) vibrational quanta.

	Now we discuss the thermal rectification effect under temperature reversal $T_L \leftrightarrow T_R$. we let $Y(\omega)=f_{L+m}(\omega)[1-f_{R-n}(\omega)]- f_{R+n}(\omega)[1-f_{L-m}(\omega)]$, and denote tildes above $Y(\omega)$ and $G_{12}^r(\omega)$ to be $\widetilde{Y}(\omega)$ and $\widetilde{G}_{12}^r(\omega)$ to be the corresponding quantities after temperature reversal. One can verify the relation $Y(\omega)=-\widetilde{Y}(\omega)=\widetilde{Y}(-\omega)$ where the equality $Y(\omega)=\widetilde{Y}(-\omega)$ is proved using $\sum_m A_{\alpha m}f_{\alpha +m}(\omega)=\sum_n A_{\alpha n}[1-f_{\alpha -n}(-\omega)]$.
	If the dressed retarded self-energy $\Sigma_{d\alpha}^r(\omega)$ is temperature independent, one can easily see that $G_{12}^r(\omega)$ in Eq.~\eqref{G12} does not change with respect to temperature reversal so that heat current is symmetric under temperature reversal. Hence the temperature dependency of dressed self-energy is the necessary condition for thermal rectification. However it is not a sufficient condition. If the junction has a symmetric coupling $\Gamma_L=\Gamma_R$, $G_{12}^r(\omega)=\widetilde{G}_{12}^r(\omega)$ holds for $\bar{\epsilon}_1=\bar{\epsilon}_2$, and $|G_{12}^r(\omega)|=|\widetilde{G}_{12}^r(-\omega)|$ holds for $\bar{\epsilon}_1=\pm\bar{\epsilon}_2$. The later is proved using parity of $\Sigma_{d\alpha}^r$ shown in Eq.~\eqref{Sr}. To conclude, in order to have thermal rectification in the DQD system with temperature dependent self-energies, one should either have asymmetric dot-lead coupling $\Gamma_L\neq \Gamma_R$ or tune the effective DQD levels such that $|\bar{\epsilon}_1| \neq |\bar{\epsilon}_2|$.

\section{Numerical Results}
In this section, we will first present the numerical calculation of the behaviors of dressed retarded self-energy with respect to temperature and electron-phonon coupling constant $g$, and then we will verify the conditions of achieving thermal rectification and examine the rectification effect by varying parameters. We assume weak coupling strengths between DQD and two leads with $\Gamma = 0.2\Gamma$ during the numerical calculations.

\subsection{Dressed retarded self-energy}
\begin{figure}
\centering
  \includegraphics[width=3.2in]{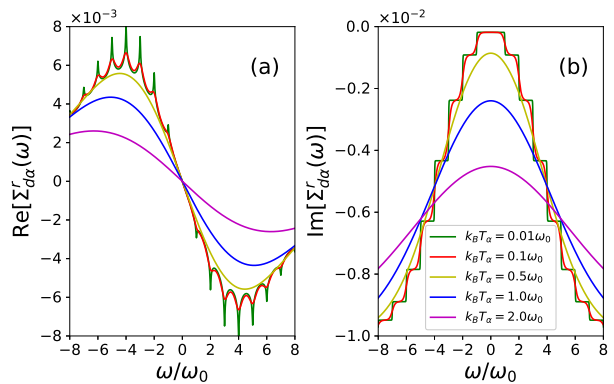} \\
  \caption{(Color online) Real and imaginary part of dressed retarded self-energies at different temperatures. Coupling constant is chosen as $g=2.0$. }
  \label{fig2}
\end{figure}

In Fig.~\ref{fig2}, we plot the real and imaginary part of dressed retarded self-energies $\Sigma^r_{d\alpha}$ at different temperatures with coupling constant $g=2.0$. We can see that the real part and imaginary part are odd and even functions of energy, respectively.
The real part of the dressed retarded self-energy shows peaks with logarithmic singularities at $n\omega_0$.\cite{BDong1}  With increasing of temperature, these peaks become small in magnitude and eventually vanish. Meanwhile, the magnitude of both real and imaginary parts decrease as well. The imaginary part has stepwise structures at low temperatures due to the opening of the inelastic channels. The stepwise structures get smoothed and vanishes at high temperatures with increasing temperatures.

\begin{figure}
  \includegraphics[width=3.4in]{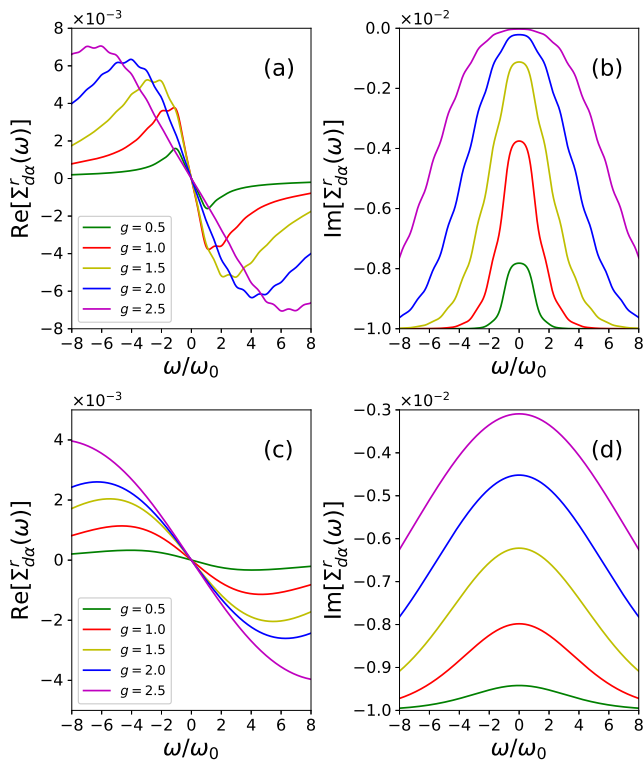} \\
  \caption{(Color online) Real and imaginary part of dressed retarded self-energies at different $g$. $k_BT_\alpha = 0.2\omega_0$ for upper panels and $k_BT_\alpha = 2.0\omega_0$ for lower panels.}
  \label{fig3}
\end{figure}

Real and imaginary part of $\Sigma^r_{d\alpha}$ at different $g$ are shown in Fig.~\ref{fig3} ($k_BT_\alpha = 0.2\omega_0$ for upper panels and $k_BT_\alpha = 2.0\omega_0$ for lower panels). With increasing of the electron-phonon coupling constant $g$, the magnitude of real part increases, and the maximal (or minimal) point shift towards smaller (larger) $\omega$. Imaginary part is always negative and increases with increasing $g$ in the whole range of $\omega$ shown.

\subsection{Thermal rectification}
In this subsection, we plot electronic heat currents $I_h$ versus $\Delta T/(2T_0)$ with $\Delta T = T_L - T_R$ and $k_B T_0= (T_L + T_R)/2 = 1.5\omega_0$ to test the condition of realizing thermal rectification. A thermal rectification ratio in this work is defined as
\begin{equation}
R_h = \frac{{\rm max}(|I_h(\Delta T>0)|, |I_h(\Delta T<0)|)}{{\rm min}(|I_h(\Delta T>0)|, |I_h(\Delta T<0)|)} -1 .
\end{equation}

\begin{figure}
  \includegraphics[width=3.4in]{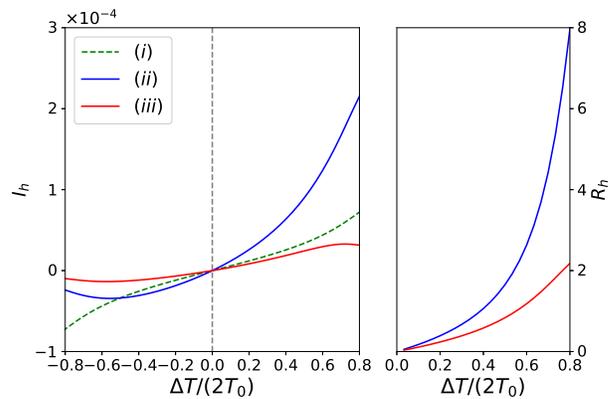} \\
  \caption{(Color online) Heat current $I_h$ as a function of $\Delta T/(2T_0)$ by varying $\bar{\epsilon}_\nu$ and $\Gamma_\alpha$: (i) $\bar{\epsilon}_1 =-\bar{\epsilon}_2= 0.3\omega_0$ with $\Gamma_L=\Gamma_R$; (ii) $\bar{\epsilon}_1 = 0$ and $\bar{\epsilon}_2 =-0.6\omega_0$ with $\Gamma_L=\Gamma_R$; and (iii) $\bar{\epsilon}_1 = \bar{\epsilon}_2 =0$ with $\Gamma_L=0.25\Gamma_R$. Other parameters are chosen as: $t_{12}=0.1\omega_0$; $g=2.0$. The corresponding rectification ratios are plotted in the right panel. }
  \label{fig4}
\end{figure}

	Heat currents versus $\Delta T/(2T_0)$ by varying effective DQD levels $\bar{\epsilon}_\nu$ and dot-lead couplings $\Gamma_\alpha$ are shown in Fig.~\ref{fig4} for the following three cases: (i) $\bar{\epsilon}_1 = -\bar{\epsilon}_2 = 0.3\omega_0$ with $\Gamma_L=\Gamma_R$; (ii) $\bar{\epsilon}_1 = 0$ and $\bar{\epsilon}_2 =-0.6\omega_0$ with $\Gamma_L=\Gamma_R$; and (iii) $\bar{\epsilon}_1 = \bar{\epsilon}_2 =0$ with $\Gamma_L=0.25\Gamma_R$. The corresponding rectification ratios are plotted in the right panel. For the case of $\bar{\epsilon}_1 = -\bar{\epsilon}_2$ with equal dot-lead coupling, heat current is symmetric with respect to temperature reversal with vanished thermal rectification ratio. Thermal rectification occurs by breaking the condition $\bar{\epsilon}_1 = \pm\bar{\epsilon}_2$ [case (ii)] or in system with unequal dot-lead couplings [case (iii)].

\begin{figure}
  \includegraphics[width=3.4in]{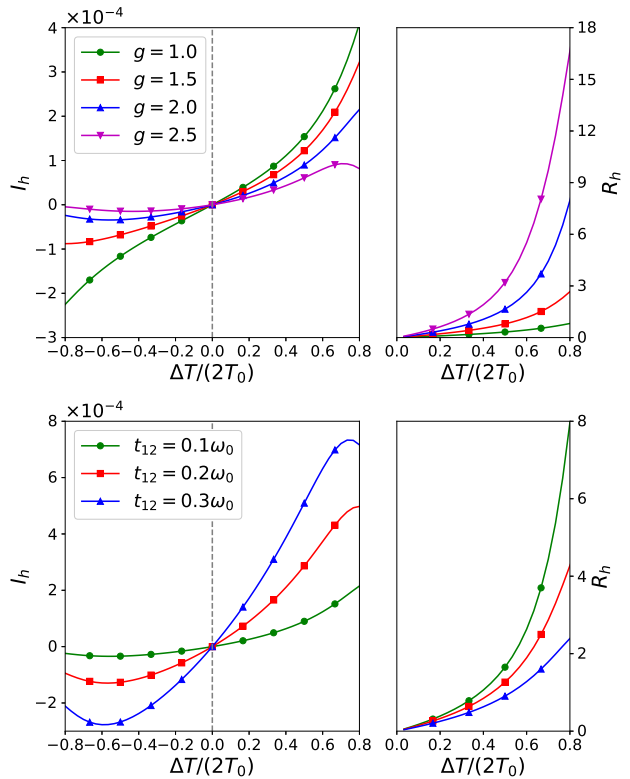} \\
  \caption{(Color online) Heat current and the corresponding heat rectification ratio as a function of $\Delta T/(2T_0)$ by varying coupling constant $g$ with $t_{12}=0.1\omega_0$ (upper panels) and inter dot hopping amplitude $t_{12}$ with $g=2.0$ (lower panels).  }
  \label{fig5}
\end{figure}

To further investigate the factors that affect thermal rectification, we show heat currents and their corresponding thermal rectification ratios for different coupling constants $g$ in the upper panels of Fig.~\ref{fig5} and inter dot hopping amplitudes $t_{12}$ in  the lower panels of Fig.~\ref{fig5}. With increasing $g$ or decreasing $t_{12}$, we observe that the heat current amplitude decreases while the thermal rectification ratio increases. This is because electron-phonon interaction strength increases with increasing $g$ so that electron becomes more difficult to escape the phonon cloud, thus reducing the thermal current amplitude. In the meantime, the nonlinearity becomes more pronounced with increasing $g$ and this is favorable to the rectification effect. Increasing inter dot hopping $t_{12}$ enables the electron to tunnel across the junction more easily and reduces the influence of electron-phonon interaction, so that heat current amplitude increases and rectification ratio decreases.

\section{Conclusion}\label{sec4}
In this work, we have studied the rectification of electronic heat current through a double quantum dot junction under a temperature bias.
The DQD in presence of strong electron-phonon interaction is described by the Anderson-Holstein model and can lead to the formation of polaronic states in which electronic states are dressed by phonon cloud.
Dressed tunneling approximation is employed to deal with strong electron-phonon interaction by dressing the self-energies with phonon cloud operator. The heat current expression is obtained using the equation of motion method. We found that the real and imaginary part of dressed retarded self-energy is, respectively, odd and even function of energy with respect to chemical potential.
The temperature dependency of dressed self-energy is due to the fact that the temperature of vibrational mode in each dot is the same with the corresponding lead it couples. This gives rise to the asymmetry of the system and is the necessary condition of thermal rectification.
On top of the temperature dependency of self-energies, one should either have asymmetric dot-lead couplings or tune DQD effective levels satisfying $|\bar{\epsilon}_1| \neq |\bar{\epsilon}_2|$ to rectify heat current.

	In the numerical calculations, we show the behaviors of dressed retarded self-energy with respect to temperatures and electron-phonon coupling constant $g$. With increasing temperature, the peaks at $n\omega_0$ of the real part of the dressed retarded self-energy become small in magnitude and eventually vanish. Meanwhile, the magnitude of both real and imaginary parts decrease, and the stepwise structures of the imaginary part get smoothed and vanishes at high temperatures. With increasing the electron-phonon coupling constant $g$, the maximal (or minimal) point shift towards smaller (larger) $\omega$. Imaginary part is always negative and increases with increasing $g$ in the whole range of $\omega$.

	The condition to realize thermal rectification by either tuning QD levels or dot-lead couplings are numerically verified.  We find that one can improve thermal rectification effect by increasing electron-phonon coupling or reducing inter dot coupling, while the electronic heat current is reduced.

\begin{acknowledgments}
This work was financially supported by the Research Grant Council (Grant No. HKU 17311116), the University Grant Council (Contract No. AoE/P-04/08) of the Government of HKSAR, NSF-China under Grant No. 11374246.
L.-Z. also thanks the financial support from the National Natural Science Foundation of China (Grant No.  11704232), National Key R\&D Program of China under Grants No. 2017YFA0304203 and No. 2016YFA0301700, Shanxi Science and Technology Department (No. 201701D121003), the Shanxi Province 100-Plan Talent Program, the Fund for Shanxi ``1331 Project" Key Subjects Construction.
\end{acknowledgments}


\end{document}